\documentclass[conference]{IEEEtran}
\IEEEoverridecommandlockouts
\usepackage{acronym}
\usepackage{amsmath,amssymb,amsfonts}
\usepackage{algorithmic}
\usepackage{cite}
\usepackage{graphicx}
\usepackage{textcomp}
\usepackage{xcolor}
\usepackage{placeins}
\usepackage{booktabs}
\usepackage{hyperref}
\usepackage{cleveref}
\usepackage{xspace}

\acrodef{acpf}[ACPF]{alternating current power flow}
\acrodef{acopf}[ACOPF]{alternating current optimal power flow}
\acrodef{scacopf}[SC ACOPF]{security-constrained alternating current optimal power flow}
\acrodef{cpu}[CPU]{central processing unit}
\acrodefplural{cpu}[CPUs]{central processing units}
\acrodef{exago}[ExaGO]{Exascale Grid Optimization Toolkit}
\acrodef{gpu}[GPU]{graphical processing unit}
\acrodefplural{gpu}[GPUs]{graphical processing units}
\acrodef{hpc}[HPC]{high performance computing}
\acrodef{mc}[MC]{marginal cost}
\acrodef{rto}[RTO]{regional transmission operator}
\acrodefplural{rto}[RTOs]{regional transmission operators}
\acrodef{uc}[UC]{unit commitment}
\acrodefplural{uc}[UCs]{unit commitments}
\acrodef{dduc}[DDUC]{data-driven unit commitment }
\acrodef{sc}[S.C.]{South Carolina}

\newcommand{\exago}{ExaGO\xspace}
\newcommand{\opflow}{OPFLOW\xspace}
\newcommand{\tcopflow}{TCOPFLOW\xspace}

\def\BibTeX{{\rm B\kern-.05em{\sc i\kern-.025em b}\kern-.08em
    T\kern-.1667em\lower.7ex\hbox{E}\kern-.125emX}}

\begin{document}

\title{Hidden Economic Consequences of Adapting to Fast Ramping Datacenter Loads\\

\thanks{This manuscript has been authored by UT-Battelle, LLC under Contract No. DE-AC05-00OR22725 with the U.S. Department of Energy. The publisher acknowledges the US government license to provide public access under the DOE Public Access Plan (http://energy.gov/downloads/doe-public-access-plan).
Willa Gutowski: wgg22@fsu.edu}
}

\author{
\IEEEauthorblockN{Willa Gutowski, Charles Foltz, Nicholson Koukpaizan,  Slaven Peles, Eve Tsybina}
\IEEEauthorblockA{Oak Ridge National Laboratory, Oak Ridge, TN, USA\\
Email: \{gutowskiwg, foltzcj1, koukpaizannk, peless, tsybinae\}@ornl.gov} }

\maketitle

\begin{abstract}
Artificial intelligence workloads are driving the rapid expansion of datacenter infrastructure, which imposes substantial stress on the US energy system. While high peak electricity prices are an anticipated outcome, measurable under peak hour simulations, the high off peak prices are a significantly underestimated threat. We simulate different ramping conditions on a congestible 5000-bus system, based on a modified IEEE 118-bus grid, to show that, in the presence of fast ramping loads and slow ramping generation, datacenters can aggravate latent load pockets. This results in unexpectedly high system costs during periods outside of datacenter peak. We test the datacenter effects using two distinct load conditions. We find that in the system coincident peak, coupled simulations result in up to 100\% loading of slow expensive units, with an average marginal cost increase of 8\%. These findings are of extreme importance as they reveal the hidden costs of preventively ramping slow generation in anticipation of datacenter load changes.
\end{abstract}


\section{Introduction}
\label{sec:introduction}

\begin{figure}[htb]
    \centering
    \includegraphics[width=0.40\linewidth]{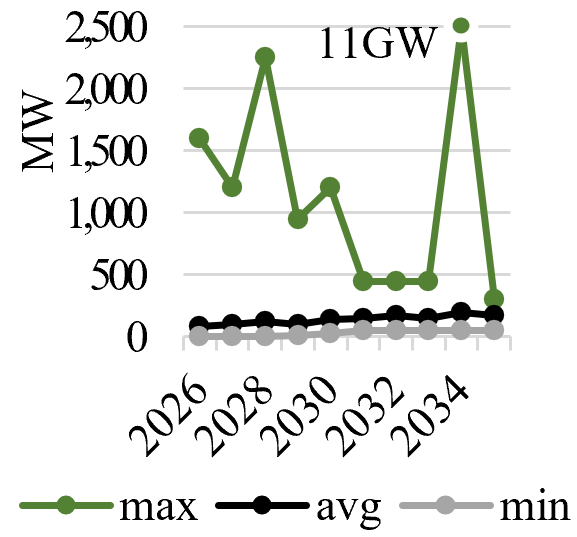}
    \hfill
    \includegraphics[width=0.58\linewidth]{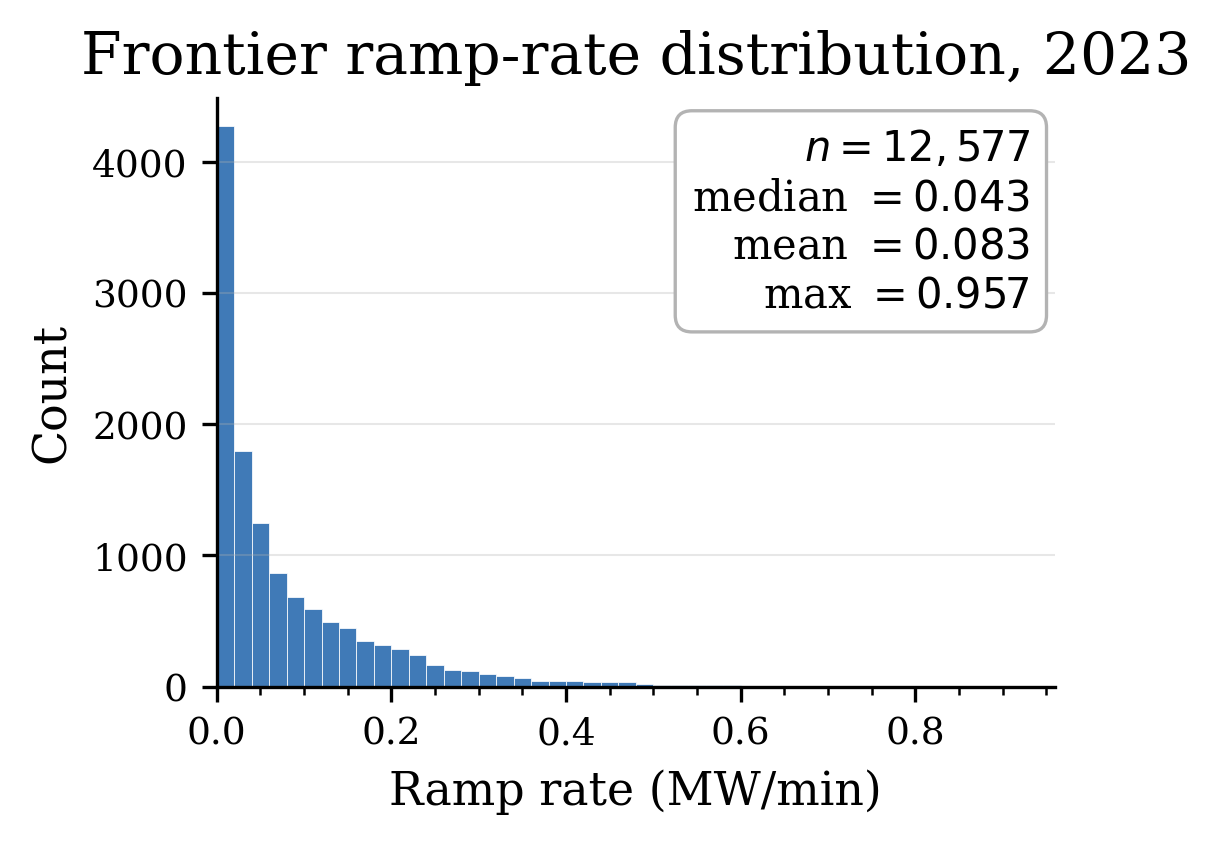}
    \caption{(a) Distribution of nominal capacity, MW, for data centers reported in \cite{SP500} as “under construction” or “land held” by year of planned entering operation. Source: \cite{SP500}. (b) Distribution of ramp time for the Frontier supercomputer over one year of observations, 2023. Ramp up starts when MW is at local minimum and increases until it is at a local maximum. Ramp down starts when MW is at local maximum and increases until it is at a local minimum. Source:\cite{Frontier}.}
    \label{fig:placeholder}
\end{figure}

There is broad consensus across industry~\cite{FERC, NERC_2026,owen_processing_2026} and academia \cite{chen2026ai_datacenters,lindberg_guide_2021,ginzburg_technical_2026,hossain_impact_2025} that the rapid expansion of datacenter infrastructure imposes substantial stress and higher costs on the US energy system. Datacenters constitute large loads that are characterized not only by their size but also by their capacity for fast ramping. Fig.~\ref{fig:placeholder}(a) demonstrates that, among the planned datacenters that report their MW capacity, this capacity can be as large as 1-2 GW. Fig.~\ref{fig:placeholder}(b) illustrates that \ac{gpu} loads can ramp at maximum rates of 1 MW/min on a 20 MW system, corresponding to as much as 5\%/min. Despite these ramping rates occurring infrequently, they must be accounted for to ensure a stable operation of the grid. While datacenter peak load has been studied relatively widely, ramping properties and their inherent challenges have received comparatively little attention. Much of the research on the multiperiod behavior of datacenters focuses on load flexibility as a grid asset~\cite{clausen_load_2014,zhang_flexibility_2020,lindberg_guide_2021,tsiligkaridis_distributed_2025}. Industry evidence, in contrast, indicates that datacenters favor firm connections~\cite{FERC} and generally do not consider shifting their loads. Taken together, these observations define a compelling space for research and discussion.

However, research cited above reveals a methodological gap that may prevent existing modeling approaches from accurately capturing intraday variations in system behavior. 
Furthermore, time-decoupled or relaxed optimization may yield misleading insights regarding which grid assets must operate to support datacenter ramping. This becomes of particular importance if the slower units are more expensive, and have to be preventively ramped up, adding high generation costs in anticipation of the datacenter load. 

We simulate different ramping conditions on a representative power grid system and contrast a simple one-hour decoupled resolution steady state analysis with a more accurate, higher-resolution multiperiod coupled approach. Our results show that once uninterrupted convergence of nodal, transmission-constrained \ac{acopf} is enforced under sharp intra-hour ramps, system behavior changes sharply. Slower units effectively become ``must run'', with substantial consequences for unit dispatch and, potentially, system cost.

The remainder of this paper is organized as follows. Section \ref{sec:design} describes the test system and outlines the simulation design. Section \ref{sec:results} presents the results along with a discussion of their implications. Section \ref{sec:conclusion} concludes with a summary of key findings and directions for future research.

\section{Simulation Design}
\label{sec:design}

We use the \ac{hpc} package \exago \cite{peles2026exago} to evaluate datacenter placement at qualifying buses in the grid. Its \opflow module provides temporally decoupled simulations, while its \tcopflow module enforces full temporal coupling. These represent the two edge cases: \opflow assesses dispatch with each period solved independently, whereas \tcopflow resolves the ramping constraints needed to evaluate the placement of fast-ramping loads such as datacenters, whose demand varies between 0 and 500~MW, consistent with the projected sizes discussed in Section \ref{sec:introduction}. \tcopflow thus reflects the most conservative, highest-reliability assumptions, and all partially coupled approaches fall between these two analyses.

\subsection{Impact Metrics}
\label{sec:proxy}

Fast-ramping large loads under transmission-constrained \ac{acopf} conditions alter unit dispatch and, potentially, total system cost. To quantify the difference between cases with and without the datacenter load, we use two proxy metrics:
\begin{itemize}
    \item the change in generator-level dispatch, in terms of (a) the MW volume and (b) the marginal cost at which those volumes are dispatched; and
    \item the change in power flow into and out of the area where the datacenter is sited, which we refer to as the \emph{import difference}.
\end{itemize}

We find the generator level dispatch by calculating the percent generator loading using actual volume $P_g$ (MW), and maximum capacity $P_{max}$ (MW). 

\begin{equation}
\mathrm{loading} = P_g / P_{max} \times 100\%
\end{equation}

The quadratic costs for fossil fuel generation would imply that it is not economically efficient to produce near $P_{max}$ where the cost of producing each additional MW of active power is increasingly high. The produced volumes are therefore likely to be far from $P_{max}$ and would be higher for relatively cheap units. If we find that a unit is loading differently for time-coupled and decoupled scenarios, this would indicate that a unit is either very low cost compared to others or ``must run'' and being dispatched for reasons other than its cost of production in a given ten-minute interval. To further narrow down the question of whether a generator is ``must run'', we assess the cost at which it produces energy similar to the locational marginal price. Assuming the cost function has the form $C(P_g) =  c + b P_g + aP_g^2$, where $a,b,c$ are constants, the marginal cost is

\begin{equation}
\left. MC = \partial C(P_g)/\partial P_g=b+2 a P_g \right|_{P_g = P_g^*}.
\end{equation}

If the unit is loaded at a higher volume and a much higher marginal cost, this would indicate that fast-ramping large loads affect unit dispatch and system cost.

We further calculate import difference between time-decoupled and fully-coupled \ac{acopf} as follows. Let $m$ represents an area in the power system and $\mathcal{T}_m$ the transmission lines adjacent to the area $m$. For each line $\ell \in \mathcal{T}_m$, $P_{\ell,m}(\tau)$ denotes the active-power flow measured at the terminal belonging to $m$. Then, the import for a module $m$ for the coupled and time-decoupled simulations are given by

\begin{equation}
I_m^{\mathrm{c}}(\tau)
=
-\sum_{\ell \in \mathcal{T}_m}
P_{\ell,m}^{\mathrm{c}}(\tau),
~~
I_m^{\mathrm{d}}(h)
=
-\sum_{\ell \in \mathcal{T}_m}
P_{\ell,m}^{\mathrm{d}}(h)
\end{equation}

$I_m(\tau)>0$ indicates that this module is a net-importer. To enable comparison, we then aggregate the fully-coupled result to hourly resolution by averaging the ten-minute periods in the set $\mathcal{S}_h$
associated with hour $h$:

\begin{equation}
\overline{I}_m^{\mathrm{c}}(h)
=
\frac{1}{|\mathcal{S}_h|}
\sum_{\tau \in \mathcal{S}_h}
I_m^{\mathrm{c}}(\tau).
\end{equation}

The import difference for each datacenter placement is then

\begin{equation}
D_m
=
\max_h
\left|
\overline{I}_m^{\mathrm{c}}(h)
-
I_m^{\mathrm{d}}(h)
\right|.
\end{equation}

We compute the metrics for the time-decoupled and fully coupled scenarios using two four-hour intervals with varying datacenter load profiles.
\paragraph{Peak demand window} The peak demand window models the datacenter load profile after the highest observed power draw. It was selected by finding the maximum demand across the dataset for the Frontier supercomputer \cite{Frontier} and centering a four-hour period around it.
\paragraph{Ramp stress or steep window} The ramp stress window models the datacenter load profile after the interval with the largest swing between maximum and minimum power draw. It was selected by finding the four-hour interval with the largest absolute difference between its maximum and minimum demand.

For each window, the decoupled simulation consists of four independent hourly optimal power flow simulations, each using the average demand for that hour; the single fully coupled simulation consists of 24 ten-minute periods; ten minutes was selected so that our ramp values used would be consistent with\cite{Frontier}. 

\subsection{Analysis Inputs}
\label{Inputs}

We simulate a transmission-constrained system based on the IEEE 118-bus model. The original model~\cite{ieee118bus} represents a near-copperplate case and lacks the complete set of properties required for testing alternative economic dispatch configurations. To adapt the original model~\cite{ieee118bus} for the present study, we introduce the following modifications, drawing primarily on the IEEE 118-bus grid operating manual~\cite{PNNL}:

\begin{itemize}
\item Making the system transmission-congestible. This requires introducing line and transformer MVA limits by appropriate voltage class and establishing transmission corridor limits on lines interconnecting areas within the larger system.
\item Introducing generator $P_{min}$, $P_{max}$, and ramping constraints.
\item Introducing generator cost. For this study, we employ quadratic cost functions for fossil-fuel units and linear cost functions for other units.
\item Making the grid scalable. We first reduce the system from 118 to 100 buses to simplify scaling, then introduce flexible scaling in increments of 100 to construct arbitrarily large systems with arbitrary area characteristics.
\end{itemize}

\begin{figure}[htb]
    \centering
    \includegraphics[width=1.0\linewidth]{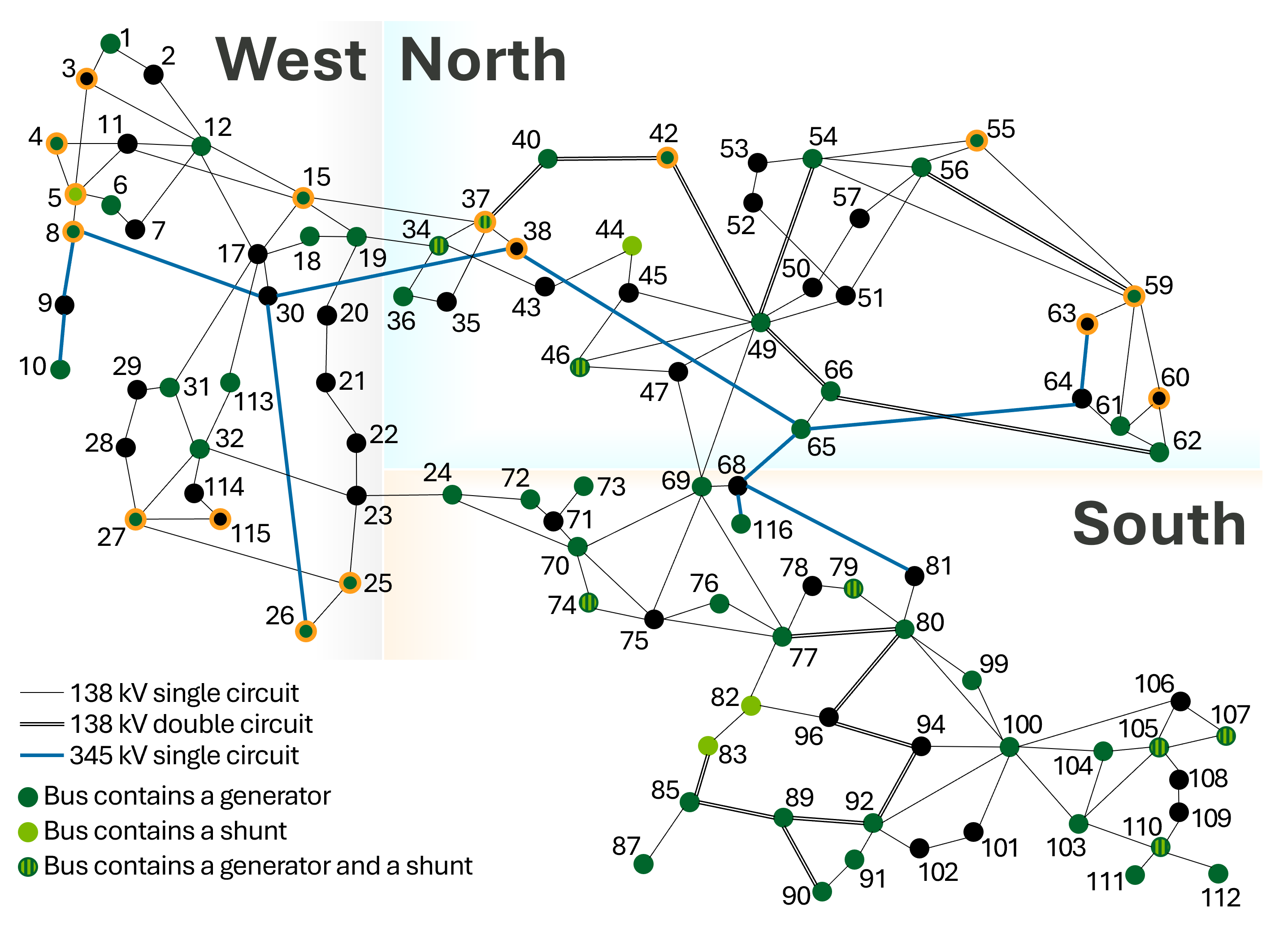}
    \caption{Modified IEEE 118-bus model to support congestion, ramping constraints, generator cost, and scaling. Yellow circles indicate buses that serve transmission corridors.}
    \label{fig:map}
\end{figure}

The resulting system is presented in Fig.~\ref{fig:map}. The connections are established by adding 118 kV lines between buses 5 $\to$ 60, 4 $\to$ 59, 3 $\to$ 55, 15 $\to$ 27, 37 $\to$ 114, and 25 $\to$ 42, while 345 kV lines connect buses 8 $\to$ 63 and buses 26 $\to$ 38.

A close examination of the modified IEEE 118-bus system shows how it allows to evaluate differences between high-resolution, fully temporally coupled studies and low-resolution decoupled or relaxed-coupled studies. We identify three characteristic areas to the system. The first (West) is higher cost but not yet transmission constrained, with slower units ramping at 10-20 MW/min. A typical example is the Appalachian \acp{rto} within PJM and MISO, whose older generation -- often small municipal coal units or large district coal units -- makes them heavily reliant on slow steam turbines. These units can also be costlier than fast natural gas generation due to rising Appalachian coal prices\cite{coal_prices}. Historically, congestion risk and slow steam generation posed no problems. Assuming datacenter loads, such areas may become costly, reliability-exposed load pockets. The West is the primary focus of our research, expected to exhibit the sharpest contrast between the traditional and the fully coupled transmission-constrained approaches.

The second area (South) is designed to be lower cost with faster-ramping units. An example is the PEPCO/PSEG areas in Pennsylvania, which host substantial local gas generation but cannot export large volumes due to a weak local grid. This area is less at risk of datacenter effects, as upfront transmission constraints make it an unlikely candidate for datacenter connection. We expect adding datacenters to the South may not lead to a feasible solution.

The third area (North) is designed to be lower cost. It also has the most connections, 50\% more than West or South. North is expected to provide balancing and wheeling benefits to adjacent areas, much as Virginia and Maryland connect other parts of the PJM grid. This area is of particular interest with respect to line MVA limits and high-resolution generation profiles.

\begin{figure}[h]
    \centering
    \includegraphics[width=0.85\columnwidth, trim={0 0 0 0}, clip]{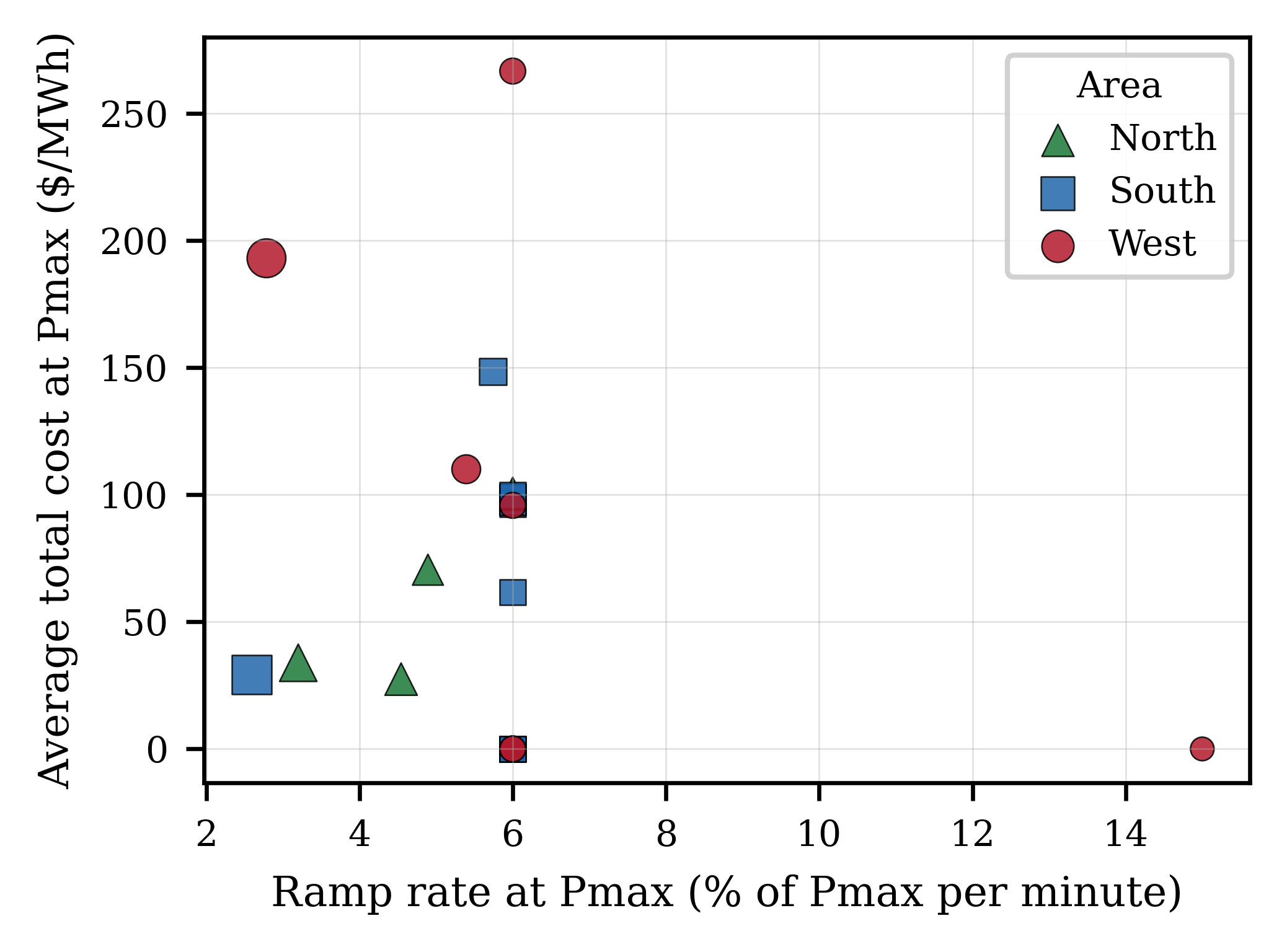}
    \caption{Generation cost and ramping by area}
    \label{fig:network_gen_us}
\end{figure}

The distribution of ramping and cost characteristics is shown in Fig.~\ref{fig:network_gen_us} for the 31 committed generating units (located at dark green buses in Fig. \ref{fig:map}) . Since the cost characteristics of units with quadratic costs are difficult to represent directly, we adopt the average total cost per MW at $P_{max}$ as a cost indicator. This is consistent with the low hot and cold reserves observed in US \acp{rto}, as datacenters tend to increase load and cause generating units to be committed more frequently and dispatched higher along the supply curve. The adopted evaluation approach shows that West has more expensive generation, whereas both North and South have cheaper generation for comparable ramp rates.

We extend the modified IEEE 118-bus system to 5000 buses, which is approximately the size of a \ac{rto}, by replicating it 50 times and interconnecting the constituent modules through transmission corridors at buses circled in yellow in Fig.~\ref{fig:map}. Connecting the three areas with more replicated systems allows for more connections, especially for West. This allows to clearly show how transmission interconnection and the resulting ability to quickly wheel power from adjacent systems affect the ability of the system to satisfy fast ramping constraints of datacenter loads. Following the expansion, 350 buses (7 buses in Fig. \ref{fig:map} connected by 345 kV lines, replicated 50 times) are capable of hosting a 500 MW datacenter. We exclude buses that have an existing generator. 

\begin{figure}[h]
\centerline{\includegraphics{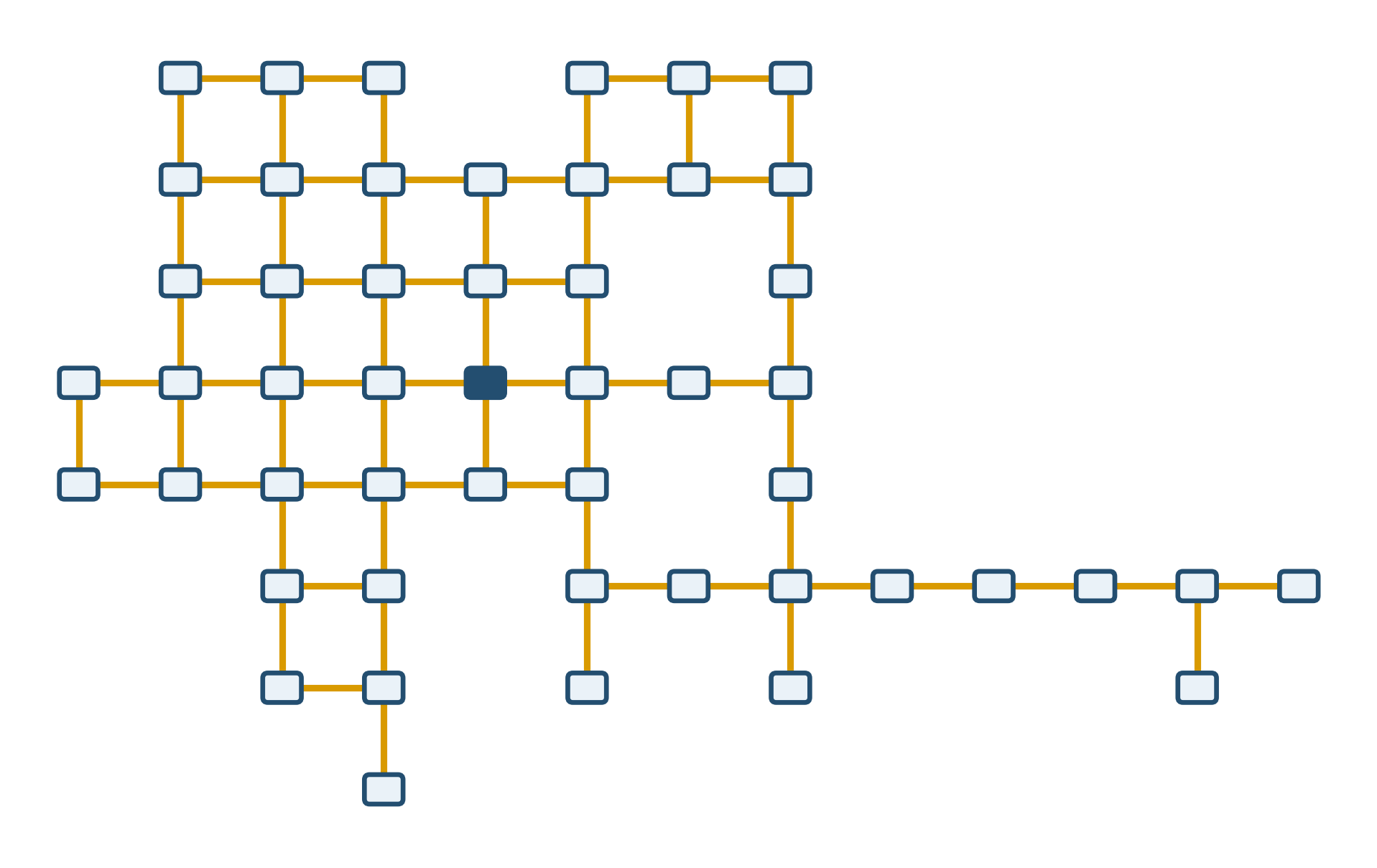}}
\caption{5,000-bus extended test system. The dark blue box denotes the original modified IEEE 118-bus system and the light boxes denote its copies (referred to as "modules"). The gold lines denote the transmission corridors connecting adjacent systems. }
\label{import}
\end{figure}

The resulting 5000-bus \ac{rto}-scale system is depicted in Fig.~\ref{import}, where each module is the modified IEEE 118-bus system. The full scaling code and the initial system properties are provided in the online appendix.

The simulations are repeated for all 350 buses capable of hosting a 500 MW datacenter. We monitor active and reactive power flows across all transmission corridors in the system and generation volumes for all committed units.


\section{Results}
\label{sec:results}

For both the peak window and the steep window, all 350 decoupled \ac{acopf} cases converged within allotted \ac{cpu} time (650 seconds, found as the minimum robust time for converged cases), while 250 of the 350 fully time-coupled cases converged. This constitutes the first important finding: all 100 time-coupled cases associated with the South area did not converge on ramping scenarios due to the local weak grid. We therefore exclude South from further analysis.

Further, we find that for the 250 converged cases, all of them in the West and North, results change depending on the location of connection of the datacenter. Placing datacenters in some locations has a stronger effect than in others. The following subsections outline in more detail the effects of datacenter placement on the observable variables of interest. 


\subsection{Results for economic dispatch}

Table \ref{tab:loading_mc} compares average generator loading percentage and average marginal cost as introduced in \ref{sec:proxy} between different areas of the modified grid. We select the average across all generators from the respective areas, as a lower bound on the loading and cost increase. Generators with steeper cost curves would show a disproportional increase in marginal cost of production. For both simulation windows, the North region had no difference (up to four-digit precision) in loading percentage or marginal cost between decoupled and time-coupled multiperiod \ac{acopf}. The high loading percentage in North is consistent with Fig.~\ref{fig:network_gen_us}, which showcases how generation is inexpensive in this region.

By contrast, in the West, loading percentage increases by 3.97 and 3.32 percentage points for peak and steep windows, respectively. Average marginal cost in this region increases by 6.87 \$/MWh for the peak window and 5.56 \$/MWh for the steep window. This lower bound estimate constitutes an increase of 7-8\% for the entire four hour window. In a real market, assuming locational marginal prices set by the single most expensive generator, the prices would increase even further (loading of the most expensive West unit doubled). This increase indicates that when fast ramping rates of data centers are factored in, more expensive generation is dispatched since expensive slow units can only provide base load under those constraints. 

The observed difference between \opflow and \tcopflow analyses indicates that ramp constraints influence dispatch decisions across time periods. The time-independent, hourly analyses select the lowest-cost feasible dispatch for each hour without accounting for the generation in subsequent periods. By contrast, \tcopflow is subject to ramp constraints that effectively cause a slow-ramping generator to maintain a higher output earlier so that sufficient generation will be available to support the fast-ramping datacenter load later.

\begin{table}[htbp]
\caption{Average Generator Loading and \ac{mc} by Area}
\label{tab:loading_mc}
\centering
\footnotesize
\begin{tabular}{|l|l|l|c|c|c|}
\hline
\textbf{Profile} & \textbf{Area} & \textbf{Method} & \textbf{Loading} & \textbf{MC} & \textbf{Cases} \\
                 &               &                 & \textbf{(\%)}    & \textbf{(\$/MWh)} & \\
\hline
\hline

Peak & North & \opflow & 100.00 & 56.96 & 150 \\
Peak & North & \tcopflow & 100.00 & 56.96 & 150 \\
\hline
Peak & West & \opflow & 79.12 & 88.35 & 100 \\
Peak & West & \tcopflow & 83.09 & 95.22 & 100 \\
\hline

Steep & North & \opflow & 100.00 & 56.96 & 150 \\
Steep & North & \tcopflow & 100.00 & 56.96 & 150 \\
\hline
Steep & West & \opflow & 79.74 & 89.75 & 100 \\
Steep & West & \tcopflow & 83.06 & 95.31 & 100 \\
\hline
\end{tabular}
\end{table}

\subsection{Results for module-import difference}

If the time coupled dispatch causes some units to behave as ``must run'', we expect to see that adding a datacenter would cause the import to be higher in decoupled case and lower in fully coupled case. We further review how the shape of the datacenter load profile impacts this behavior.

\begin{figure}[htb]
\centering
\includegraphics[width=0.8\columnwidth]{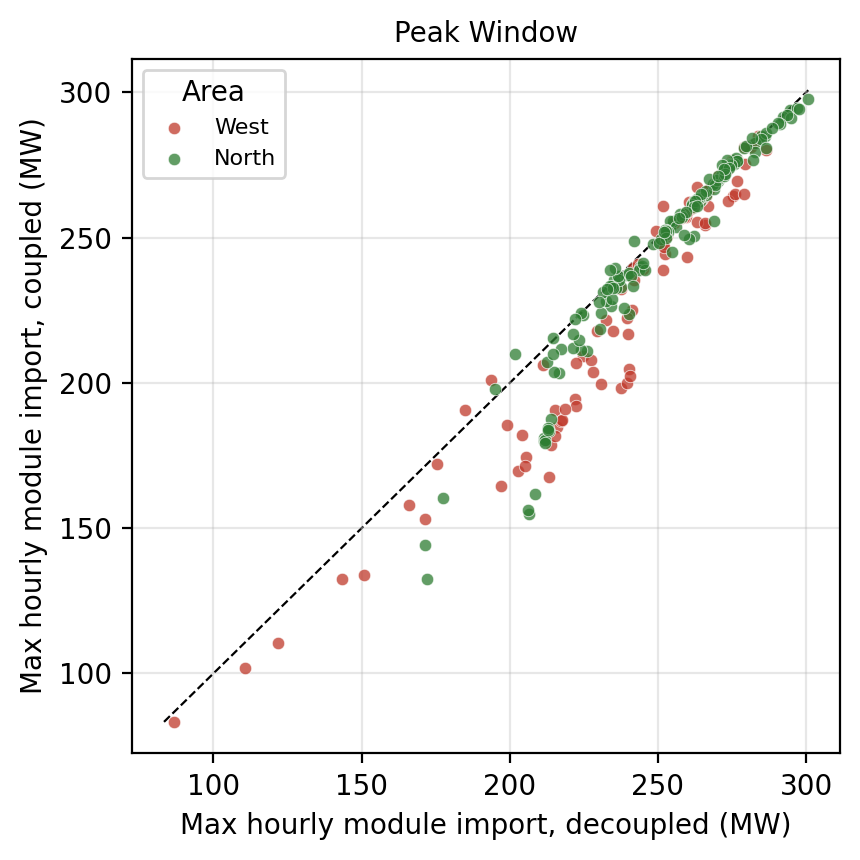}
\includegraphics[width=0.8\columnwidth]{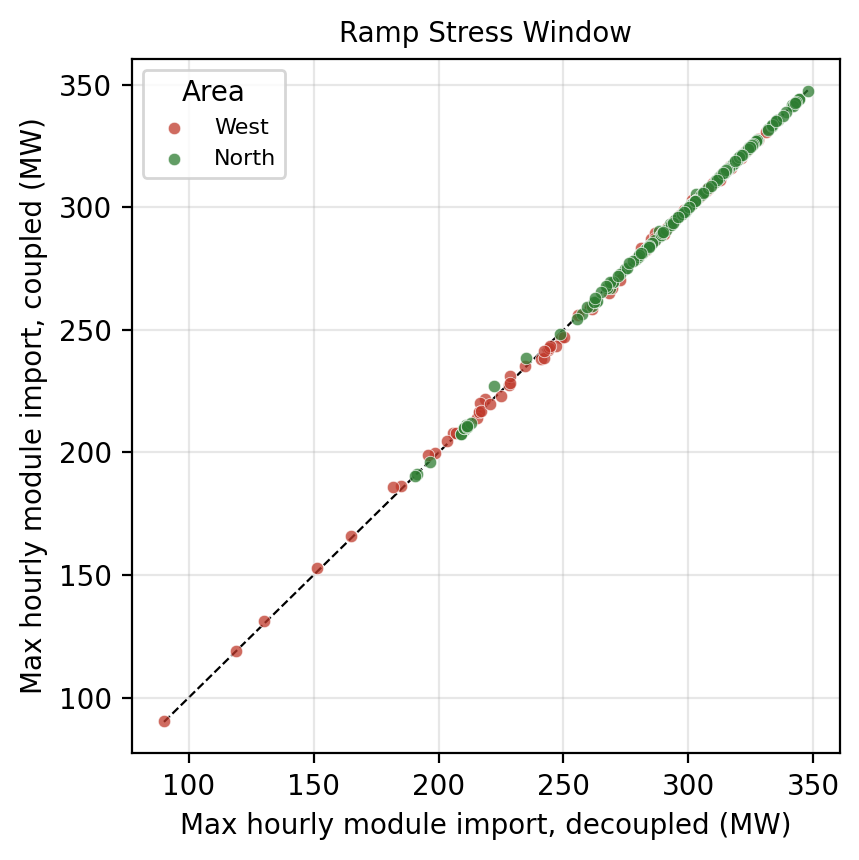}
\label{West_import}
\caption{Comparison of fully decoupled and fully coupled module import values for each candidate datacenter location. The color indicates the area of datacenter placement.}
\label{greendots}
\end{figure}

Figure~\ref{greendots} 
shows that, for the peak scenario, West imports up to 15-20\% more in time-decoupled analysis, while North has some outliers but the two analyses produce mostly similar results. This confirms the findings above: the slow and expensive ``must run'' units force the West system to produce locally in coupled scenario, but are used less in the decoupled scenario. For the ramp stress scenario, there is no difference between coupled and decoupled simulations, indicating that datacenters do not affect dispatch. 

Overall results show that the peak window ($r = 0.965$) has larger deviations from this line than the ramp-stress window ($r = 0.999$). Moreover, the results presented in the peak window are consistent with our hypothesis that the West ($r = 0.963$) would exhibit greater deviations between coupled and decoupled simulations than the North ($r = 0.968$). The difference only occurs in the peak window, which indicates that the way in which the datacenter load is applied impacts potential divergence between coupled and decoupled simulations. The difference in imports mostly happens in the West region, which proves the argument that in potential load pockets the expensive generation becomes ``must run'' reducing the need for imports.



\FloatBarrier

\section{Conclusions}
\label{sec:conclusion}

This paper presents a comparison of coupled and decoupled simulations of a congestible 5000-bus system, based on a modified IEEE 118-bus grid, when a datacenter load is applied to candidate buses. 

Firstly, we find that system impact, as measured by loading percentage and marginal cost, varies by generator location within the grid. Generators in the West region had the largest differences in these metrics between coupled and decoupled scenarios on average, indicating that coupled simulations cause more expensive generation to be dispatched because of the ``must run'' effect observed in slow-ramping units. 

Secondly, we examine differences in module import between coupled and decoupled simulations. We similarly find the largest differences within the West region, indicating that in potential load pockets, expensive generation becomes ``must run'' and the need for imports is reduced. In future work, we plan to investigate  battery storage as a potential strategy to mitigate excess dispatch of slow-ramping generators.  

\section*{Acknowledgments}
This research used resources of the Oak Ridge Leadership Computing Facility at the Oak Ridge National Laboratory, which is supported by the Advanced Scientific Computing Research programs in the Office of Science of the U.S. Department of Energy under Contract No. DE-AC05-00OR22725.

\bibliographystyle{IEEEtran}
\bibliography{references}

\end{document}